\documentstyle[12pt,epsfig]{cernart}
\def\la{\mathrel{\mathpalette\fun <}}
\def\ga{\mathrel{\mathpalette\fun >}}
\def\fun#1#2{\lower3.6pt\vbox{\baselineskip0pt\lineskip.9pt
\ialign{$\mathsurround=0pt#1\hfil##\hfil$\crcr#2\crcr\sim\crcr}}}

\newcommand{\bc}{\begin{center}}
\newcommand{\ec}{\end{center}}
\newcommand{\bd}{\begin{displaymath}}
\newcommand{\ed}{\end{displaymath}}
\newcommand{\be}{\begin{equation}}
\newcommand{\ee}{\end{equation}}
\newcommand{\ba}{\begin{array}}
\newcommand{\ea}{\end{array}}
\newcommand{\bt}{\begin{tabular}}
\newcommand{\et}{\end{tabular}}

\begin{document}
\large
~
\vspace{1cm}

\bc

{\Large\bf On Importance of Inelastic Rescatterings\\[0.6cm] in
Pion Double Charge Exchange on Nuclei}\\[1cm]
{\Large A.B.Kaidalov, A.P.Krutenkova}\\[1cm]
{\Large\it State Research Center \\
Institute of Theoretical and Experimental Physics \\
Moscow, 117259, Russia}\\[1cm]
\ec
\begin{abstract}
Different aspects of the Gribov--Glauber  approach for calculation of 
inclusive pion double charge exchange (DCX) on nuclei are investigated.
Recently we have shown that inelastic rescatterings (IR) with two 
(and more) pions in the intermediate states give an important contribution 
to the process of DCX at 
energies above $\sim0.6\;GeV$. In this paper 
we use the one pion exchange model
to study in details amplitudes of two pion production. This
allows us to verify theoretical assumptions made in the previous paper
and to predict cross section for the forward inclusive pion
DCX at energies $\ga1\;GeV$, where
IR dominate over the conventional DCX mechanism
of two sequential single charge exchanges.

\end{abstract}

PACS:\\
Keywords:\\

\vspace{0.5cm}
\section{Introduction}
\smallskip
The inclusive pion double charge exchange (DCX) reaction on nuclei, e.g.\\
\be
\pi^- + A(Z,N)\rightarrow \pi^+ + X 
\ee   
as a two-step transition on two like nucleons (protons) is described by the 
diagrams of three types (Fig.1(a), (b), (c)) which correspond to the
Glauber picture of elastic (a), quasielastic (b) and inelastic (c)
rescatterings.

The conventional DCX mechanism of two sequential single charge exchanges
(SSCX), i.e. the elastic rescattering, is found to describe [1],  in
general, experimental data on pion  DCX at energies of the 
meson factories up to 0.5 $GeV$. It predicts [2] a 
strong decrease of forward angle exclusive pion DCX at incident 
kinetic energies $T_0\ga0.6\;GeV$ due to the rapid energy drop 
of $\pi^-\rightarrow\pi^0$ transition rate.
The SSCX mechanism leads [3] to
analogous behaviour of inclusive forward DCX rate too.

Inclusive reaction (1) is a unique process where, 
as it was shown in [4], the effect of the Glauber inelastic rescatterings
(IR) becomes very important already at $GeV$ energies.
This effect can be expected
from a comparison of experimental 
total cross sections for two competing processes, $\pi^-p\rightarrow\pi^0n $
and $\pi^-p\rightarrow\pi^+\pi^-n $ (Fig.2,
where data are given from the compilation [5]). The study of the
forward angle reaction (1) by ITEP experiments [6] on $^{16}O$ and
$^{6}Li$ in a kinematics which forbids real production of additional pion
has demonstrated a relatively weak energy dependence 
of its rate and considerable
excess over SSCX mechanism prediction [6] around
$T_0 = 1\;GeV$. So the conventional elastic 
SSCX picture of inclusive pion DCX has
to be modified by an important new mechanism of IR at $ GeV$ energies.

Our goal is to elaborate recently suggested Gribov-Glauber approach 
to inclusive pion DCX reaction [4] on $^{16}O$.
Below we present an updated analysis of some theoretical aspects of our model
and  clarify  simplifying assumptions we  have used.
Then we estimate quantitatively upper and lower limits for inclusive
pion DCX cross section in the energy region of $T_0 = 1~ GeV$ to 4$~ GeV$
to be tested in future experiments. Our working scheme for the calculation of
$\pi\rightarrow 2\pi$ amplitude is the one pion exchange (OPE) model
which we check by a comparison of  theoretical  cross section with experiment
as well as distributions on two-pion effective mass
for the process $\pi^-p\rightarrow\pi^+\pi^-n $. We compare
our results with  predictions of other models and in particular
of the meson exchange currents (MEC) model [7].

\section{Gribov-Glauber picture of inclusive pion DCX}
\smallskip
In this Section we shall discuss some aspects of the formalism for
a calculation of the amplitude of the DCX process $\pi^- pp \to \pi^+ nn$
(Fig.3). Following Gribov [8] it is convenient to introduce an integration
over $M^2_H$, square of mass of the intermediate state in Fig.1.
It is related to integration over Fermi motion of initial protons
and momenta of final neutrons. These momenta are limited
by experimental conditions. For example in the ITEP experiment
on $^6Li$
and $^{16}O$ [6] the inclusive DCX reaction $A(\pi^-,\pi^+)X$ was measured
at kinetic energies
$T_0=0.6$, $0.75$ and $1.1\;GeV$  ($\langle\theta\rangle\approx5^0$) with
kinematical condition  $\Delta T=T_0-T\le m_{\pi}$ 
($T$ is the kinetic energy of the outgoing pion) in order to exclude
additional pion production.

The kinematical region indicated above strongly limits momentum transfers
$t_{1(2)}$ to neutrons in the transitions $\pi^-p\rightarrow Hn$
 and $Hp \to \pi^+n$ (see Fig.3).
$M_H$ can be very large at high incident energy, $E_0$, and increases as
$M^2_H\sim2m_NE_0$.
The value of $M^2_H$ varies also due to the Fermi motion.
 It is easy to obtain an estimate of this variation
taking into account that at high energies, $E_{0}\gg{m_{N}}$,
$|t_{1(2)}^{min}|$ $\approx {(M^2_H-m_{\pi}^2)}^2/4E_0^2$
$\la{p_F}^2\la2m_N\Delta T_{1(2)}$
and $(M_H^2)^{max}\approx 2$$E_0(2\Delta T_im_N)^{1/2}$.
For energy $E_0\sim1\;GeV$, $(M_H^2)^{max}$ is equal to $0.5\;GeV^2$.
As it was explained in Ref. [4], the integral over $M^2_H$ can be written
either over the real axis or (due to analyticity in  $M^2_H$) as a sum of the
absorptive part and integral over semicircle at $|M^2_H|$ $\sim$ 
($M^2_H)^{max}$.
At high enough energies $E_0$ the last contribution can be neglected if the
corresponding amplitude decreases faster than $1/M^2_H$ at large  $M^2_H$.
We gave arguments [4] that the amplitude 
of Fig.3 should satisfy this property.
In this case the DCX amplitude is proportional to the discontinuity in  $M^2_H$
of the amplitude in Fig.3, and using unitarity we obtain the representation
equivalent to the diagrams of Fig.1.

Contribution of a single particle intermediate state ($\pi^0, \eta^0$) or
narrow resonance ($\omega,...$) to the DCX cross section can be written
in the form

\be
\frac {d\sigma^{IR}_{DCX}}{d\Omega} \propto \left.(
\int\int \frac{d^{2} \sigma_{\pi \rightarrow H}}
{dt dM^{2}} dt dM^{2} \right.)^{2},~~~   H = \pi^0, \eta^0, \omega,...  
\ee

Let us comment on contribution to DCX of different diagrams of Fig.1
that was taken in Ref.[4] additively. The diagram ($b$) gives
a small correction to the Glauber elastic rescattering of Fig.1($a$)
due to the fact that $\pi^-\to\eta$ amplitude is relatively small
(see Fig.2) while the production of two pions is a very important
competing mechanism, especially for  $T_0\ga0.6\;GeV$.
As for diagrams  ($a$) and ($c$), if $\pi\to2\pi$ amplitude is dominated
by the pion exchange (see Fig.4), then their interference is absent.
Really, the invariant form of the nucleon vertex in Fig.4 is
$\bar u\gamma_5u$, while the amplitude of the transition 
$\pi^-p\to\pi^0n$ is proportional to $\bar u(A+B\hat q)u$, where 
$q = p_{\pi^-} + p_{\pi^0}$. So the interference term is 
Tr$(\hat p_1 + m_N)\gamma _5(\hat p'_1 + m_N)(A+B\hat q)$ = 0.
 
Note that for the $s$-wave $\pi\pi$ production which is a dominant
inelastic process at energies $E_0\la1\;GeV$ the interference is absent
by the same reason.
For the $s$ wave (and in general for all even waves) the same expression (2)
is valid. However for realistic case of $\pi \pi$ production, when both
 even and odd orbital momenta are important, the amplitude of $\pi^- \to
\pi^+$ transition (contrary to the diagonal
$\pi^- \to \pi^-$ transition) can not be written [4] in the form (2).
This is especially clear in the $\pi$ exchange  (OPE) model (Fig.4) where
DCX amplitude is expressed in terms of the $\pi^-\pi^+\to \pi^+\pi^-$
amplitude (or backward elastic $\pi^-\pi^+$ scattering amplitude).

In this paper we shall use the OPE model in order to study the problems
outlined above: is it possible to neglect the contribution
of the large semicircle at energies $~$ 1$~GeV$ and what is the difference
between integrals of forward and backward $\pi^-\pi^+$ scattering amplitudes?
In this way we shall find corrections $\Gamma_H$ to simple expressions
for DCX cross sections 
\be
\frac{d\sigma_{DCX}}{d\Omega}=
\frac{d\sigma^{\pi^0}_{DCX}}{d\Omega}\left(1+\sum_H\Gamma_H \right)\\
\ee 
where
\be
\left.
\Gamma_{\pi^+\pi^-}=\left(\int dM
\int\limits^{t^{\exp}_{1\max}}_{t_{1\min}(M)}
\frac{d^2\sigma_{\pi \to \pi^+\pi^-}(M,t_1)}{dMdt_1}dt_1\right/
\int\limits^{t^{\exp}_{1\max}}_{0}(d\sigma_{\pi^0}/dt_1)dt_1\right)^2,
\ee
\be
\left.
\Gamma_{\pi^0 \pi^0}=\left(\sigma^{tot}_{\pi \to \pi^0\pi^0}\right/
\sigma^{tot}_{\pi \to \pi^+\pi^-}\right)^2\cdot\Gamma_{\pi^+\pi^-}
\ee
introduced in Ref.[4] and shall obtain lower and upper bounds for these cross
sections at higher energies.

\section{Inelastic rescatterings and comparison with ITEP experiment}
\smallskip
In this Section we shall calculate 
the contribution of IR to DCX using Eqs.(3)-(5) and
experimental data on particle production in reactions \\
\be
\pi^-p \rightarrow Hn,~~~ H = \pi^0,~~~ \pi^+ \pi^-,~~~
 \pi^0 \pi^0, ~~.~~.~~. \\
\ee

The quantities $\Gamma_H$ in Eqs.(3)-(5) 
were determined using experimental data
for the following hadronic
states $H$: $\pi^0$ [9], $\pi^+\pi^-$ and $\pi^0 \pi^0$ [10].
While integrating over $t$ and $M^2$, experimental constraints on 
$\Delta T$ are taken into account (see details in [4]): 
$|t_{1(2)}|\le 2m_N(\Delta T^{max}/2)$
[\footnote{In principle it is necessary to integrate over
$t_1$ and  $t_2$, but taking into consideration some uncertainty of the
theoretical estimates as well as of the experimental information
available we limited ourselves with this
simple relation.}].

In the Table 1 we show the updated summary of energy
dependence for forward inclusive DCX reaction cross sections of $\pi^-$
on $^{16}O$ obtained in experiments at ITEP [6]
and calculated via Eqs.(3)-(5).
The first line in the Table 1 contains the values of
$d\sigma^{\pi^0}_{DCX}/d\Omega$ calculated [3,6] in the framework
of the elastic SSCX mechanism without taking into account 
effect of the Fermi motion
(FM) on magnitude of the $\pi N$-interaction energy in the nucleus.
In the values $d\tilde\sigma^{\pi^0}_{DCX}/d\Omega$ of the second
line  the FM effect
\footnote{We thank L.Alvarez-Ruso and M.J.Vicente Vacas 
for preparing the version
of the code [3] which takes into consideration the Fermi motion.}
is included. The 
resulting values are $d\sigma_{DCX}/d\Omega$ = $d\tilde
\sigma^{\pi^0}_{DCX}/d\Omega$ + ($\Gamma_{\pi^+\pi^-}+\Gamma_{\pi^0\pi^0})
$ $d\sigma^{\pi^0}_{DCX}/d\Omega$ are given at the fourth line.
Due to a sharp change of DCX rate 
at energies of interest the FM effect
gives smearing of the dip-bump structure, 
that in average leads in some increase
of the values of DCX cross section.
Note, that all values in the Table 1
correspond to the outgoing momentum spectra integrated over the
region of $\Delta T$ = 0 to 140 $ MeV $.\\

\section{Inclusive pion DCX in the energy range of 1 -- 4 $ GeV$}
\smallskip
We see from the experimental data compilation of Fig.2 that the cross section
values of the reaction (6) with $H = \pi^+ \pi^-$,\\

$~~~~~~~~~~~~~~~~~~$ $~ \pi^-p \rightarrow \pi^+ \pi^-n,~$ $~~~~~~~~~~~~~~~
$~$~~~~~~~~~~~~~(6')$\\
\\
exceed ones with  $H \neq \pi^+ \pi^- $ not only at 1 $GeV$ but
also at higher energies. So it is reasonable to assume that IR with two pion
intermediate states will dominate inclusive pion DCX at least up to
4$~ GeV$. In this case the diagram of Fig.4 can be used to obtain
a quantitative estimate  of DCX rate within the OPE model. 
Note here that, besides of the well-known good description of experiment for
(6$'$) reaction, the OPE model is very convenient to our goals because of its 
factorization feature. This permits to separate $M^2$ and $t$ dependences
of the (6$'$) reaction amplitude, $ A_{\pi\to2\pi}(M^2,t)$, and to take into
 account limitations on their values in the experiment [6].

We start from the following formula \\
\be
\frac{d\sigma^{\pi \pi}_{DCX}}{d\Omega} \propto
\left(
\int\limits^{M^2(t_{max})}_{4m^2_\pi} dM^2\int\limits^{t_{\max}}
_{t_{\min}(M^2)} dt~ \mbox{\rm Im}~ A^{forward}_{\pi\to 2\pi}
(M^2,t)\right)^2 \\
\ee
equivalent to Eqs.(3)-(5).
Then we use the OPE model [11]
to calculate a difference between imaginary parts
of forward and backward amplitudes and to estimate the
real part of the amplitude.

In OPE model it is possible to express the amplitude of
the process (6$'$) in terms of 
the $\pi\pi$ scattering amplitude, \\
\be
\mbox{\rm }~ A_{\pi\to2\pi}(M^2,t) = \tilde F^2(t)~ \mbox{\rm }
A_{\pi\pi}(M^2)\\
\ee 
where 
\be
\tilde F^2(t) = 2 G^{2}
~ [|t|/(t - m^2_\pi)^2]~ \mbox{\rm exp}[2R^{2}(t- m^2_\pi)],~~R^{2}~=~1.92 
~GeV^{-2},
\ee
$G$ is the $\pi NN$-coupling constant. \\
\noindent
$(a)~~OPE~ model~ for~ DCX$

As it was shown in [11]
the OPE model calculations are in a good agreement with the experimental data
on the reaction (6$'$) above ~2 $ GeV$. So for calculation
of the cross section for this reaction we shall follow Ref.[11],
 namely:\\
\be
\frac {d^{2}\sigma_{\pi \rightarrow 2\pi}}{dM^{2} dt}~=~G^{2}
\frac{Q(M^{2},m^2_\pi,m^2_\pi)M }
{2^{4}\pi^{2}Q^{2}(s,m^2, m^2_\pi)s} \tilde F^2(t) ~
\sigma^{tot}_{\pi \pi}(M^{2})\\ 
\ee
 where $G^{2}/4\pi=$14.6,~ 
$Q(s,m_{1}^{2},m_{2}^{2})$ =
$\sqrt{s^{2}-2s(m_{1}^{2}+m_{2}^{2})+(m_{1}^{2}-m_{2}^{2})}/2\sqrt{s}$, \\[3mm]
$M$ is the mass of two-pion state, \\
\be
\sigma_{\pi^{+} \pi^{-}}(M^{2})~=~2 \pi \int 
\mid f_{\pi^{+} \pi^{-}} \mid ^{2} dz,~~z = cos\theta, \\
\ee
 and   the $\pi\pi$ scattering amplitude is equal to
\be
f_{\pi \pi}(M^{2},z)~ =
~\frac{1}{Q(M^{2})}
\sum_{l}(2l+1)[1+(-1)^{I+l}]f_{l}^{I}(M^{2}) P_{l}(z). \\
\ee
 By virtue of the unitarity condition we have:
\be
\mbox{\rm Im}A_{\pi \pi}(M^2,z=1) = 2Q(M^2,m^2_{\pi},m^2_{\pi})
M\sigma^{tot}_{\pi\pi}(M^2). \\
\ee 
The partial-wave amplitudes $f^I_l(M^2,z)$ 
for orbital angular momentum $l$ and isospin $I$ 
are expressed via the phase shifts, $\delta_{l}^{I}$, and elasticities,
 $\eta^I_l$:\\
\be
f_{l}^{I}(M^{2})~ =~ \frac{1}{2i}[\eta^{I}_{l}(M^{2})
e^{2i\delta_{l}^{I}(M^{2})}-1 ],~~  f_{\pi^{+} \pi^{-}}~=~
\frac{1}{6}f^{2} + \frac{1}{3}f^{0} + \frac{1}{2}f^{1}.\\
\ee

It should be noted that the  equation (7) takes into account both  
$\pi^+\pi^-$ and  $\pi^0\pi^0$ intermediate states.

In the following we will use the complete set of amplitudes
(phase shifts and elasticities) with $l$ = 0, 2 for $I$ = 0, 2 and with
$l$ = 1, 3 for $I$ = 1 in the range of $M$ up to $\approx$1.8 $GeV$
from Ref.[12] where fixed-$t$ and fixed-$u$ analyticity in conjunction
with energy independent phase-shift analysis gave the single solution.
The following features of the model confirm
that its predictions are reasonable. \\
(1) The $M^2$ dependence of the total cross section of $\pi^+\pi^-$scattering
calculated according to Eq.(11), (13), (14) is in a good agreement with
the data from Ref.[13] (see Fig.5).\\
(2) The value of the parameter $R^2$ = 1.918 $GeV^{-2}$ that we use was
obtained on the base of our analysis  of the $t$
dependence of the
yield for the reaction (6$'$) averaged over the energy region from
2 to 3 $GeV$ taken from [11].
The total cross section of this reaction calculated via the integration of 
Eq.(10),
\be
\sigma_{\pi^- p \to \pi^+\pi^- n}=
\int\limits^{M^2(t_{max})}_{4m^2_{\pi}} dM^2
\int\limits^{t_{\max}}_{t_{\min}(M^2)} dt
\frac{d^2 \sigma_{\pi^- p \to \pi^+\pi^- n}}{dM^2dt} \\
\ee
(for $t_{\max}$ = $t_{max}(s))$ is compared to experimental data 
($full ~~ stars$) from Ref.[5]
in Fig.6. It is seen that the model used ($full~~ circles$) reproduces
experimental energy behaviour of $\sigma_{\pi^- p \to \pi^+ \pi^- n}$ only
above  $\approx$2.5 $GeV$. Some underestimate of the absolute value
of the cross section decreases  as $T_0$ increases.

In Fig.6 the cross section of the reaction (6$'$),
obtained with constraints of 
$t_{max} = t^{exp}_{max} = 0.135~ (GeV/c)^2$
corresponding to the experimental limitation $\Delta T \leq 140~ MeV$,
is also presented. The experimental values ($crosses$ and $empty ~~ star$) 
were calculated from the data
of Ref.[10a] on $M^2$ and angular cross-section dependencies of
the reaction (6$'$) ($crosses$) and taken from Ref.[14] for
the interval $1.5 \leq t/m^{2}_{\pi} \leq 8$ ($empty ~~ star$).
The OPE model used ($triangles$ in Fig.6) 
agrees with experiment starting already from 1 $GeV$.
Note that below $T_0 = 1~ GeV$ the OPE model predictions strongly
deviate from experimental data. This is probably 
related to a substantial role 
of $s$-channel resonance production in this region.\\
(3) The experimental distributions on the two-pion mass ($M$ or $M^2$) for
incident momentum  values of 1.343 [10], 1.59 [14], 2.26 [15], and 4
$GeV/c$ [16] are
compared  with the calculated ones (see Figs. 7-10). It is seen
from Fig.7 that the 
model predicts too small cross section in the
low $M^2$ region at the lowest
energy. On the other hand, at higher
energies the OPE model reproduces experiment rather well (including $\rho$
and $f_2(1270)$ resonances as it is seen in Figs. 9, 10). For our purpose
it is essential to have a good description
of experiment in $t$ interval close to $t \leq 0.135~(GeV/c)^{2}$. 
It follows from Fig.8 that for the region of 
$1.5 \leq t/m^{2}_{\pi} \leq 8$ an agreement 
with experiment is better than for the 
total cross section of (6$'$).

Thus we conclude that it is reasonable to use the OPE model 
for  calculation of  the high-energy DCX cross section.\\

\noindent
$b)~~Tests ~of~ the~ assumptions$

Now it is interesting to use the OPE  model calculations for checking
of the simplifying assumptions of our approach to DCX formulated in
Ref.[4] and discussed above.

The real and imaginary parts of forward ($z$ = 1) and backward ($z = -1$)
invariant amplitude, $A_{\pi^+ \pi^-}(M^2,z)$, were calculated 
using Eqs.(12) -- (14). 
As can be seen from Fig.11, Re$A_{\pi^+ \pi^-}(M^2,z=-1)$ oscillates but it
is not small compared to Im$A_{\pi^+ \pi^-}(M^2,z=-1)$
so its contribution should be taken into account at energies
~ $\sim1~GeV$. 
This means that an integral over  Re$A_{\pi^+ \pi^-}(M^2,z=-1)$
can not be neglected in this energy range and thus the same is
true for integral over the large semicircle. 
The backward parts of the amplitude appeared to be rather
 different from the forward ones within various intervals of $M^2$.
The actual ranges of $M^2$ for a set of initial energies 
for experimental data avaliable on the reaction (6$'$)
are given in the Table 2.
 
In the Fig.12 we present the energy dependence of the
cross section 
$d\sigma^{\pi \pi}_{DCX}/d\Omega$ calculated according
Eq.(7) at 
$M^{2}_{max}|_{140}$  
for the following cases:\\
(1) Im$A^{forward}_{\pi \to 2\pi}$ is obtained through the Eq.(8) for the
forward ($z = 1$) amplitudes;\\
(2) Im$A^{forward}_{\pi \to 2\pi}$ is replaced by Im$A^{backward}_{\pi \to
2\pi}$ which is obtained according to Eq.(8) for the imaginary part
of the backward ($z = -1$) $\pi \pi$ amplitude;\\
(3) Same as in (2), but for Re$A^{backward}_{\pi \to
2\pi}$;\\
(4) both imaginary and real parts of the backward
amplitude are taken into account.

The above analysis shows that for $\pi^+\pi^-$-scattering amplitudes there 
are substantial differencies between integrals of forward and backward 
scattering amplitudes and they can be neglected only at rather low
energies, where $s$-wave production dominates. We think however that OPE 
model overestimates this difference, especially for  $T_0\la1\;GeV$, as it
predicts too low $s$-wave production at these energies (see Fig.7).
So the OPE-result for DCX can be considered as a lower bound for this
cross section.

\section{Results}
\smallskip
Let us now calculate the cross section for the reaction\\
\be
\pi^-~ +~  ^{16}O \to \pi^+~ +~ X
\ee
in the kinematical range $0\le \Delta T \le 140~MeV$ of the experiment [6]
as a sum of the cross sections
with $\pi^0$ and 2$\pi$ in intermediate states (see Eq.(3) ),
\be
\langle
\frac{d\sigma_{DCX}}{d\Omega}\rangle _{140} =
\langle \frac{d\tilde\sigma^{\pi^0}_{DCX}}{d\Omega}\rangle_{140}
+ \langle \frac{d\sigma^
{\pi \pi}_{DCX}}{d\Omega}\rangle_{140}.\\
\ee 
Here $\langle d\tilde\sigma^{\pi^0}_{DCX}/ d\Omega \rangle_{140}$
is the cross section of the reaction (16) within the SSCX mechanism
(FM effect is taken into account) 
which is shown in Fig.13 (see $dashed$ curve).
$\langle d\sigma^
{\pi \pi}_{DCX}/d\Omega \rangle_{140}$ is the
contribution of the intermediate $\pi \pi$ state
for the given kinematical range of $\Delta T$.
For Im$A^{forward}_{\pi \to
2\pi}$ according to Eqs.(3)-(5) and normalizing at 
$T_{o}~=~0.6~GeV$ we obtain the $dotted~curve$ in Fig.13.
The solid curve corresponds to an account of the
full backward scattering amplitude. It can be considered
as a lower bound for the DCX cross section, while
the dotted curve can be considered as an upper
bound for the cross section. 
Experimental point at 1.1 $GeV$ is close to this bound.
This is related in our opinion to an underestimate of the $s$-wave
contribution in the OPE model at these energies noted above. 
So at higher $T_0$
we expect future DCX measurements to be closer to the solid curve.
Note, that both curves at $T_{o} < 1~GeV$, where
OPE model strongly deviates from experimental data
(see Fig.6), are calculated using not theoretical
but experimental values (see the first two 
$crosses$ in Fig.6). 

The contribution from three-pion intermediate state
to Eq.(17) estimated using the cross section of
$\pi^- p \rightarrow \omega n$ appeared to be
not more than 10\% even at the highest energy
considered.  

\section{Comparison with other approaches}
\smallskip
Let us discuss a relation between inelastic rescatterings considered
in this paper, and other approaches which have been proposed 
for the DCX process.

Relation of the present approach to a model of 
meson exchanged currents (MEC) [18]
has been discussed in Ref.[4]. The main difference is that in the OPE model
we treat the $\pi^-\pi^+$-scattering amplitude as a function of $M^2$
and integrate over this variable while in 
the MEC model the $\pi^-\pi^+$-amplitude
is approximated as a point-like interaction taken at the threshold
($M = 2m_{\pi}$). Due to soft pion theorems this amplitude is small and
thus leads to small modifications of SSCX predictions [7]. We take into
account in the OPE model both real and imaginary parts of $\pi^-\pi^+$
amplitude in the regions of $M^2$ where they are not small, so it is not
surprising that our results for DCX cross sections are substantially higher
than SSCX predictions (even for a solid curve in Fig. 13).

Another possible mechanism of DCX [19] is a production of an intermediate
$\Delta$ isobar on one proton of a nucleus, $\pi^- p\to \pi^+\Delta^-$,
with its absorption [20]
on a second proton, $\Delta^- p \to nn$. A characteristic
feature of this mechanism is a fast decrease of DCX cross section at 
$T_0 > 1~GeV$ due to energy dependence of the $\sigma(\pi^- p \to \pi^+
\Delta^-)$ as is shown in Fig.2 by the curve. 

Our study shows that at least up to energies $\sim 4~GeV$ the region of
$M^2_H \sim M^2_{max}$ is very essential in integrals over $M^2_H$, and
integrals of both imaginary and real parts of amplitudes are important. 
This means that average distances, $d$, between participating nucleons are
relatively small: $d \sim (2m_N\Delta T)^{-1/2}$. These nucleons are closely
correlated and can be even considered as a 6-quark system 
(as, for example, in Ref.[21]).

\section{Experimental outlook and conclusions}
\smallskip
We have extended the conventional Glauber mechanism for DCX process with
an account of inelastic rescatterings. In this paper we investigated
the problems which arise due to a non-diagonal nature of the process
using the OPE model. This model gives a reasonable description of the
reaction $\pi^- p \to \pi^-\pi^+ n$ at energies above $2~GeV$. So we
think that our predictions for DCX process based on this model, which
are substantially higher than SSCX results, are reliable in this energy
region. DCX is not yet studied experimentally at energies above
$T_0 = 1.1~GeV$. If strong deviations from our predictions will be found
it will have important implications: either other mechanisms 
(e.g. those mentioned
in the previous Section) are important or the phase-shift analysis of $\pi \pi$
scattering [12], which we have used, is not reliable in the large-mass region.
An interesting information on these problems can be obtained by varying
experimental interval of $\Delta T$. Note that for small $\Delta T$ (and in
particular in the exclusive limit) an integration region in $M^2$ is small,
and results should be closer to SSCX predictions.

Thus experimental study of the energy 
dependence of DCX in the region of $T_0$ =$1\div5\; GeV$ can give 
an important information on dynamics of this process.
Cross section in this region according to our estimate is not too small,
 and experiments seems feasible.
  However, at higher energies the possibility to
measure the cross section  in the region of
$\Delta T=0\div140\;MeV$ will be limited by the energy resolution 
of the detector.

In conclusion, it is important to stress once more that the
investigation of the pion DCX reactions in the intermediate energy
region can provide a test on a validity of the Glauber--Gribov approach
 in the case when effects of inelastic rescatterings are large.
 Another class of
processes where such inelastic contributions are very important is
heavy ions interactions at very high energies. It is known (see, for
example, [22]) that these contributions strongly 
modify a usual Glauber picture and an equilibration 
in the system can appear only as a result of such
rescatterings effects. 

\section*{Acknowledgments}
\smallskip
A.P.K. thanks 
I.S. and I.I.Tsukerman for 
permanent 
support. This work 
was supported in part by RFBR Grant No. 98-02-17179.

\clearpage
\bigskip
\bt{l|lll}
$T_0,\;GeV$                    &0.6&0.75&1.12\\[0.15cm]\hline
 & & &\\
$d\sigma^{\pi^0}_{DCX}/d\Omega,\,\mu b/sr~~ [8]$&
125.0&10.4&3.1\\[0.25cm]
$d\tilde\sigma^{\pi^0}_{DCX}/d\Omega,\,\mu b/sr~~
$&
139.3&25.4&4.7\\[0.25cm]
$\Gamma_{\pi^+\pi^-}+\Gamma_{\pi^0\pi^0}$&0.11&1.75
&11.3\\[0.25cm]
$d\sigma_{DCX}/d\Omega,\,\mu b/sr$&153.1
&43.6&39.7\\[0.25cm] \hline
 & & &\\
$d\sigma^{\exp}_{DCX}/d\Omega,\,\mu b/sr~~[6]$
&$59.6\pm7.4$&$43.3\pm5.5$&$26.6\pm8.9$\\[-0.15cm]
\\
\hline
\et
\vspace{0.5cm}
\begin{center}
Table 1.
\end{center}

\bigskip
\vspace*{1.0cm}
\bt{l|l|l}
$T_0,\;GeV$ &$M^2_{max}(s)$&$M^2_{max}(s)|_{140}$\\[0.15cm]\hline
 & & \\
0.61&0.36&0.36\\[0.25cm]
0.88&0.57&0.51\\[0.25cm]
1.02&0.69&0.60\\[0.25cm]
1.20&0.84&0.69\\[0.25cm]
1.45&1.08&0.84\\[0.25cm]
1.76&1.74&1.02\\[0.25cm]
2.12&2.02&1.26\\[0.25cm]   
2.86&2.5&1.68\\[0.25cm]
3.86&3.2&2.2\\
[0cm]\\
\hline
\et
\\
\vspace*{0.5cm}

\begin{center}
Table 2.
\end{center}

\clearpage

\begin{figure}[bhtp]
\begin{center}
\vspace*{-3.0cm}
\mbox{\epsfig
{figure=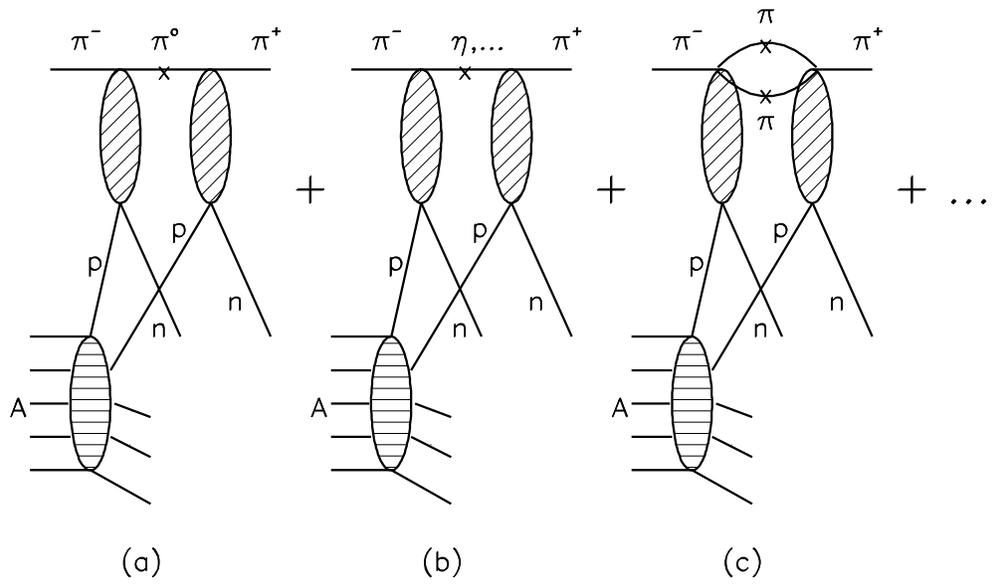,width=17cm,
            bbllx=0,bblly=0,bburx=567,bbury=841}}
\caption [] {Diagrams contributing to pion double charge exchange
on nuclei:

(a) sequential single charge exchanges (SSCX), i.e. standard
mechanism (elastic rescattering),

(b) quasielastic rescatterings,

(c) inelastic rescatterings.}
\label{diagram1}
\end{center}
\end{figure}

\clearpage
\begin{figure}[bhtp]
\begin{center}
\vspace*{-3.0cm}
\mbox{\epsfig
{figure=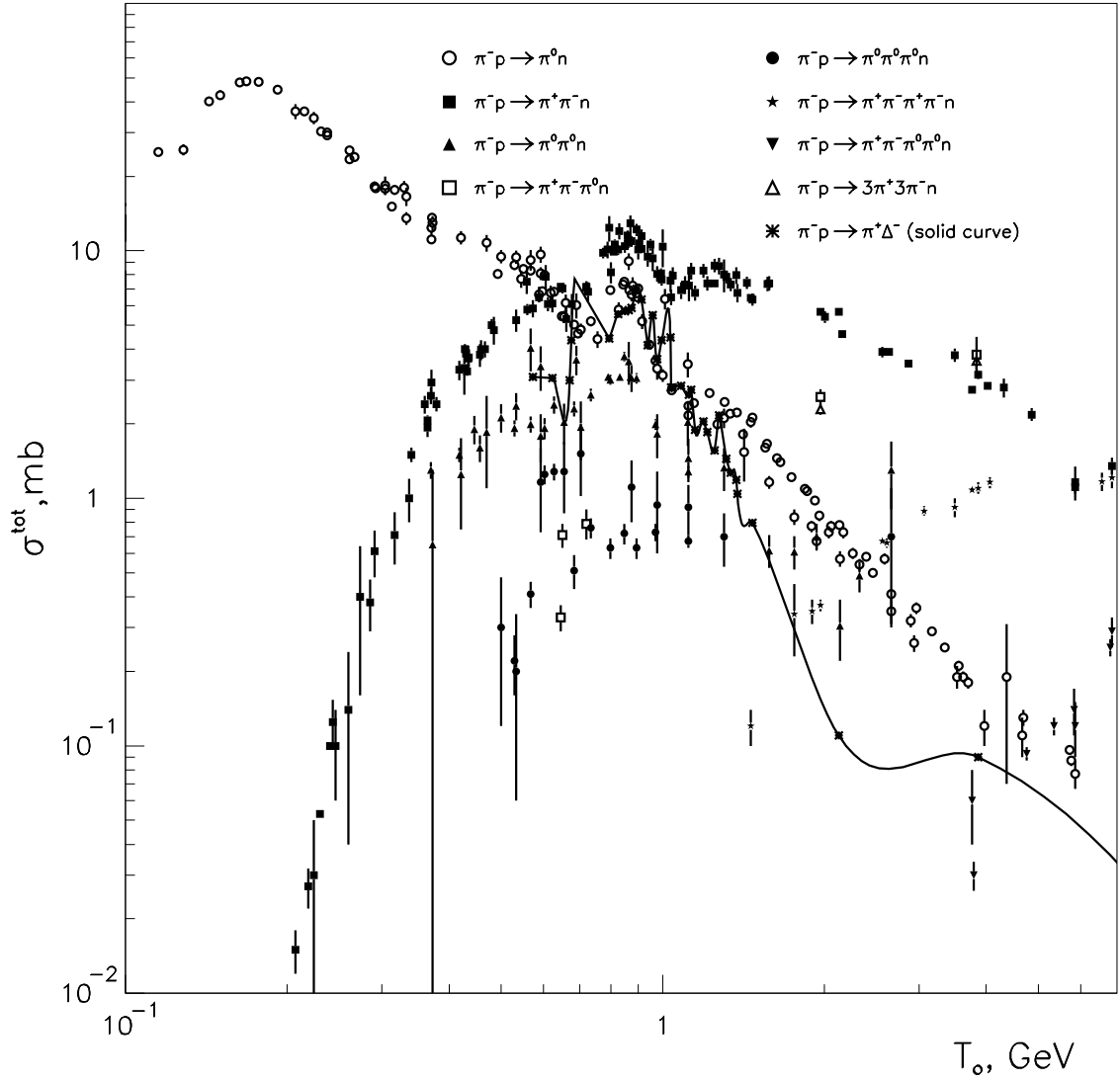,width=17cm,
            bbllx=0,bblly=0,bburx=567,bbury=841}}
\caption [] {Experimental total cross section of the reaction 
$\pi^{-}p \rightarrow Hn$ and $\pi^{-}p \rightarrow
\pi^{+} \Delta^{-}$ (see compilation [5]).}
\label{diagram2}
\end{center}
\end{figure}

\clearpage
\begin{figure}[bhtp]
\begin{center}
\vspace*{-3.0cm}
\mbox{\epsfig
{figure=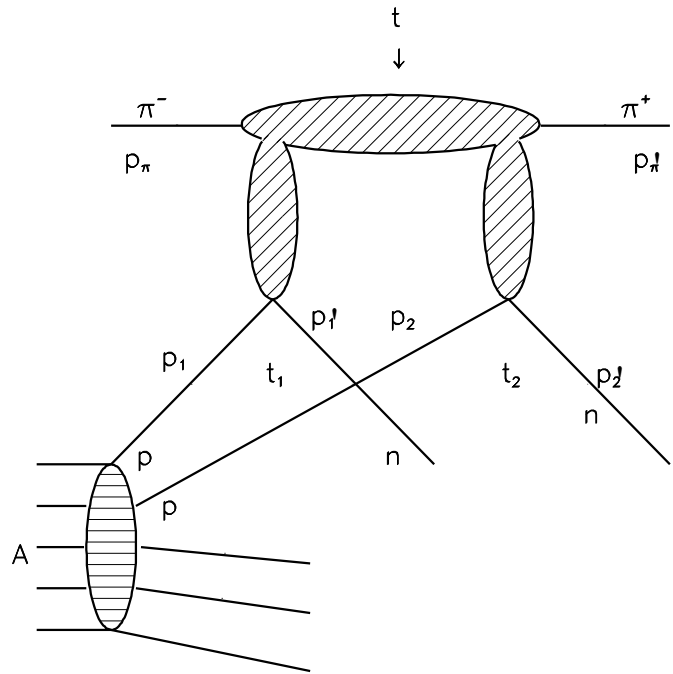,width=17cm,
            bbllx=0,bblly=0,bburx=567,bbury=841}}
\caption [] {Pion double charge exchange on nucleus
(the most general diagram).}
\label{diagram3}
\end{center}
\end{figure}

\clearpage
\begin{figure}[bhtp]
\begin{center}
\vspace*{-5.0cm}
\mbox{\epsfig
{figure=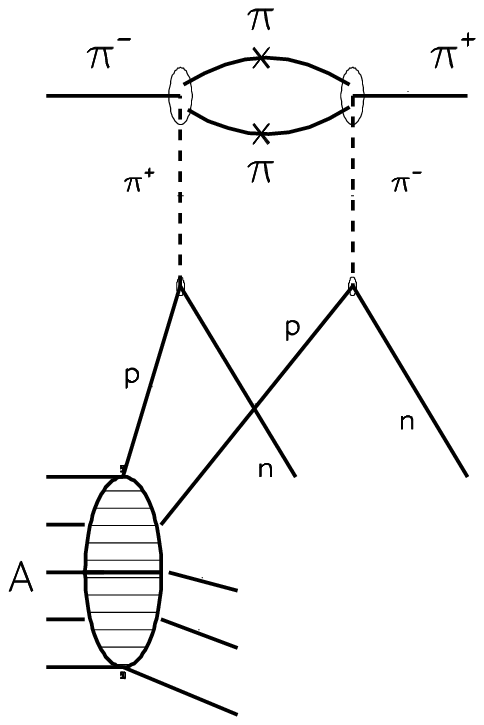,width=20cm,
            bbllx=0,bblly=0,bburx=567,bbury=841}}
\caption [] {Pion double charge exchange in one pion exchange model.}
\label{diagram4}
\end{center}
\end{figure}

\clearpage
\begin{figure}[bhtp]
\begin{center}
\vspace*{-3.0cm}
\mbox{\epsfig
{figure=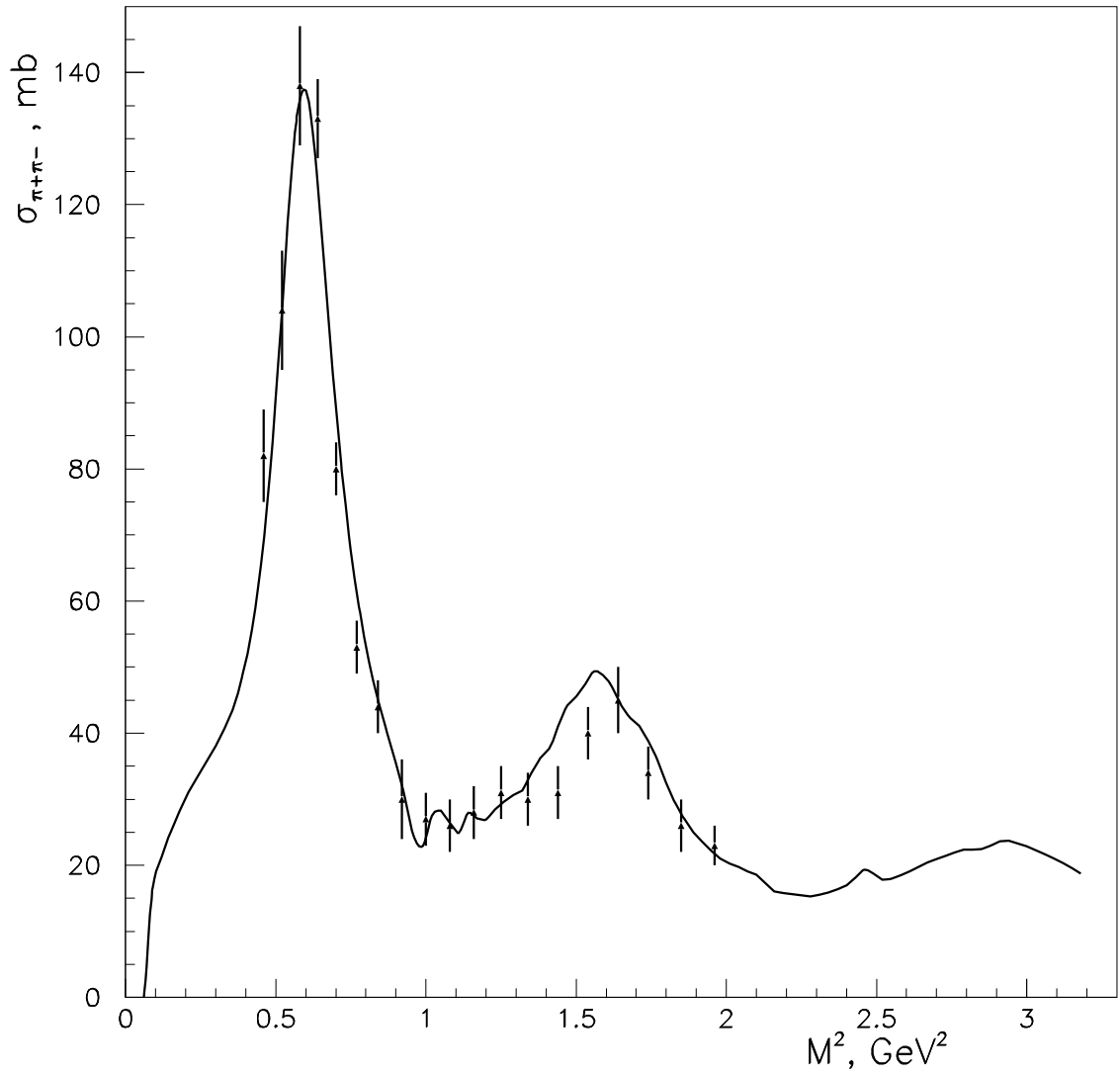,width=17cm,
            bbllx=0,bblly=0,bburx=567,bbury=841}}
\caption [] {Total cross section of the reaction
$\pi^{+} \pi^{-} \rightarrow \pi^{+} \pi^{-}$ 
from [13]; the solid curve is our calculation.}
\label{diagram5}
\end{center}
\end{figure}

\clearpage
\begin{figure}[bhtp]
\begin{center}
\vspace*{-3.0cm}
\mbox{\epsfig
{figure=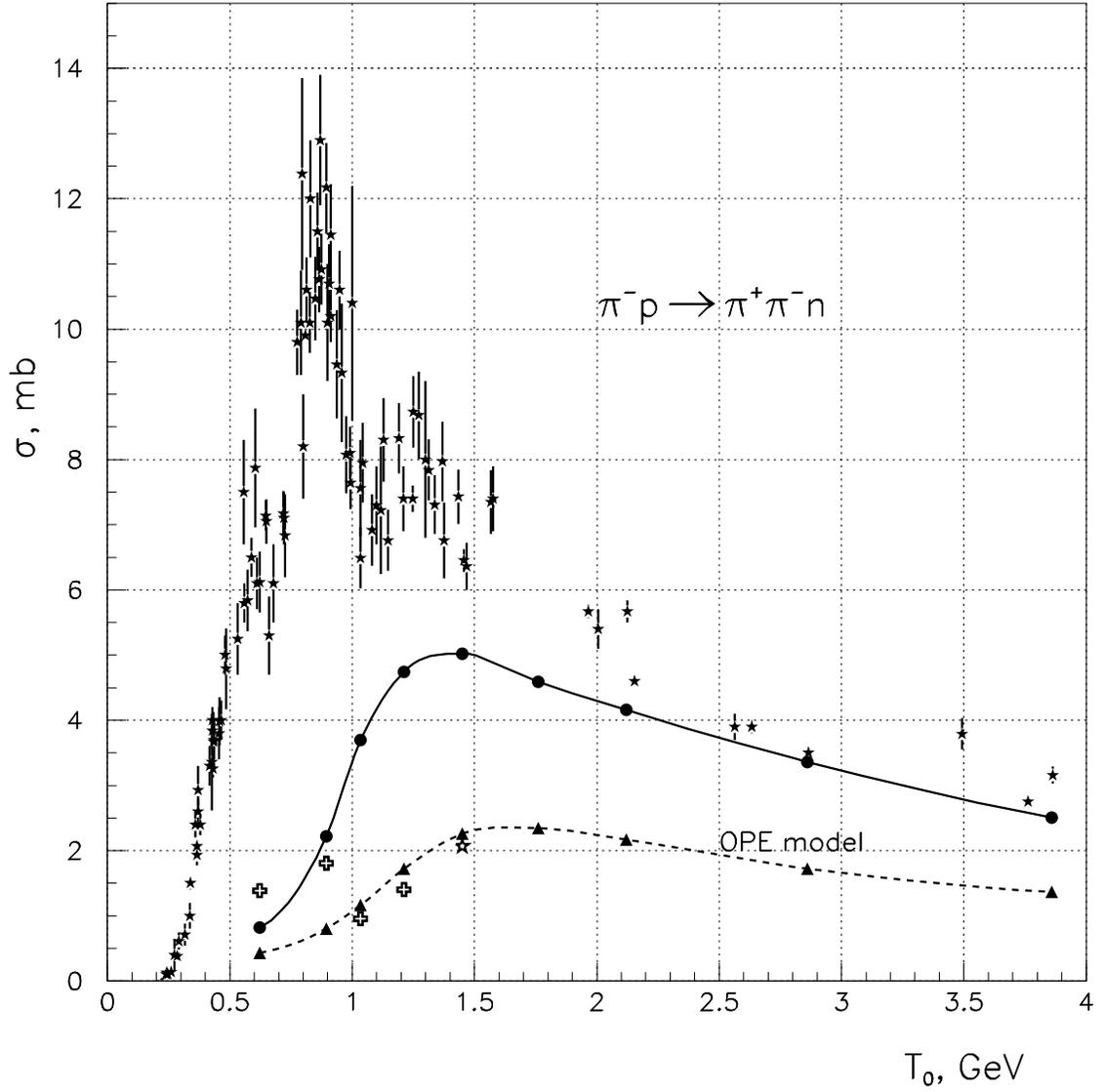,width=17cm,
            bbllx=0,bblly=0,bburx=567,bbury=841}}
\caption [] {Total cross section of the reaction
$\pi^{-}p \rightarrow \pi^{+} \pi^{-}n$. 
Stars and circles are experimental [5] and OPE
cross sections, crosses and triangles are their
corresponding truncated values for the $\Delta T \leq 140~MeV$;
empty star is for limited $t$ (see text).}

\label{diagram6}
\end{center}
\end{figure}

\clearpage
\begin{figure}[bhtp]
\begin{center}
\vspace*{-3.0cm}
\mbox{\epsfig
{figure=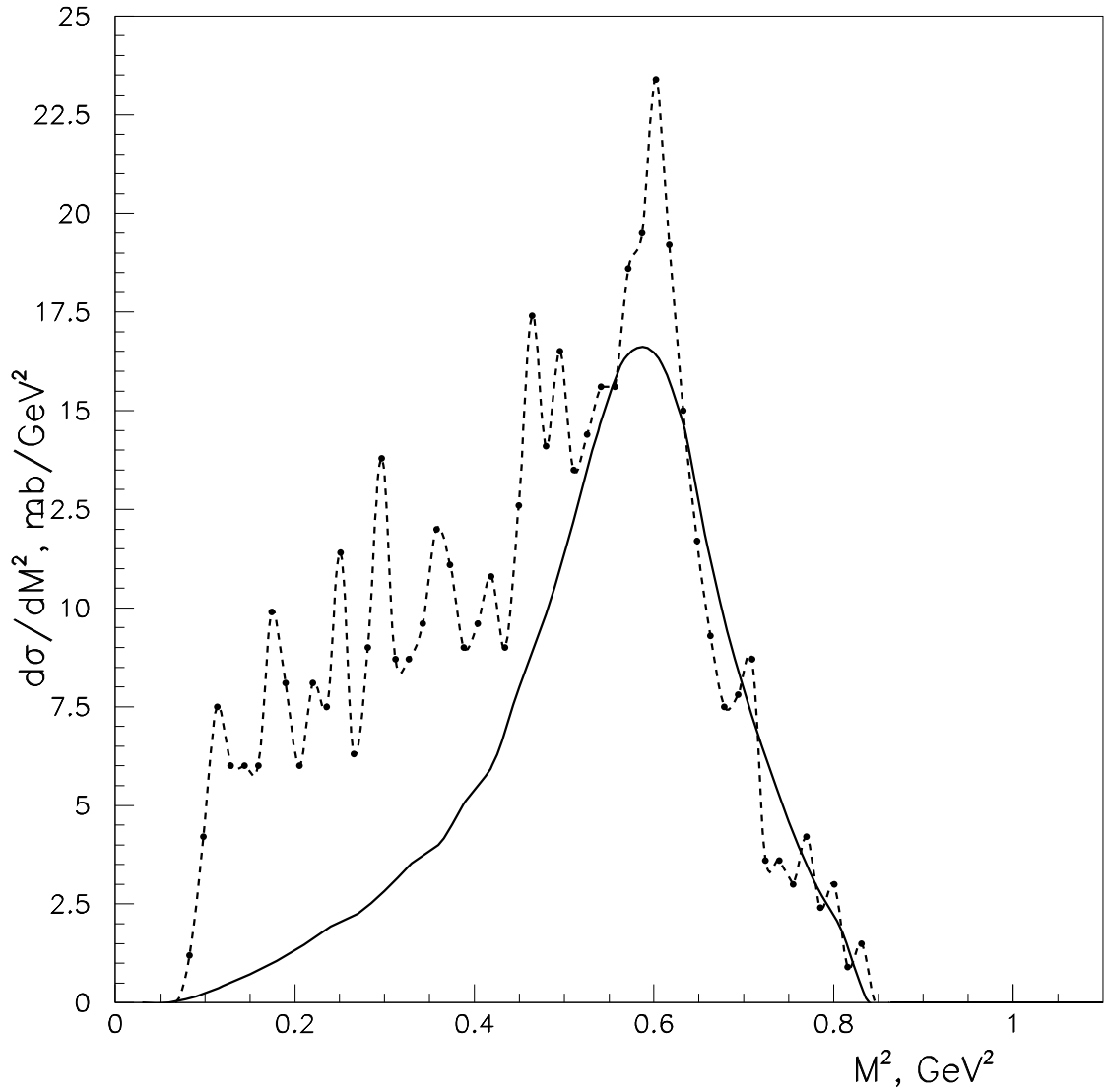,width=17cm,
            bbllx=0,bblly=0,bburx=567,bbury=841}}
\caption [] {Mass-squared distribution of 
$\pi^{+} \pi^{-}$ system in the reaction 
$\pi^{-}p \rightarrow \pi^{+} \pi^{-} n$ [10a]
at $p = 1.343~GeV/c$.
The solid curve is our calculation.}
\label{diagram7}
\end{center}
\end{figure}

\clearpage
\begin{figure}[bhtp]
\begin{center}
\vspace*{-3.0cm}
\mbox{\epsfig
{figure=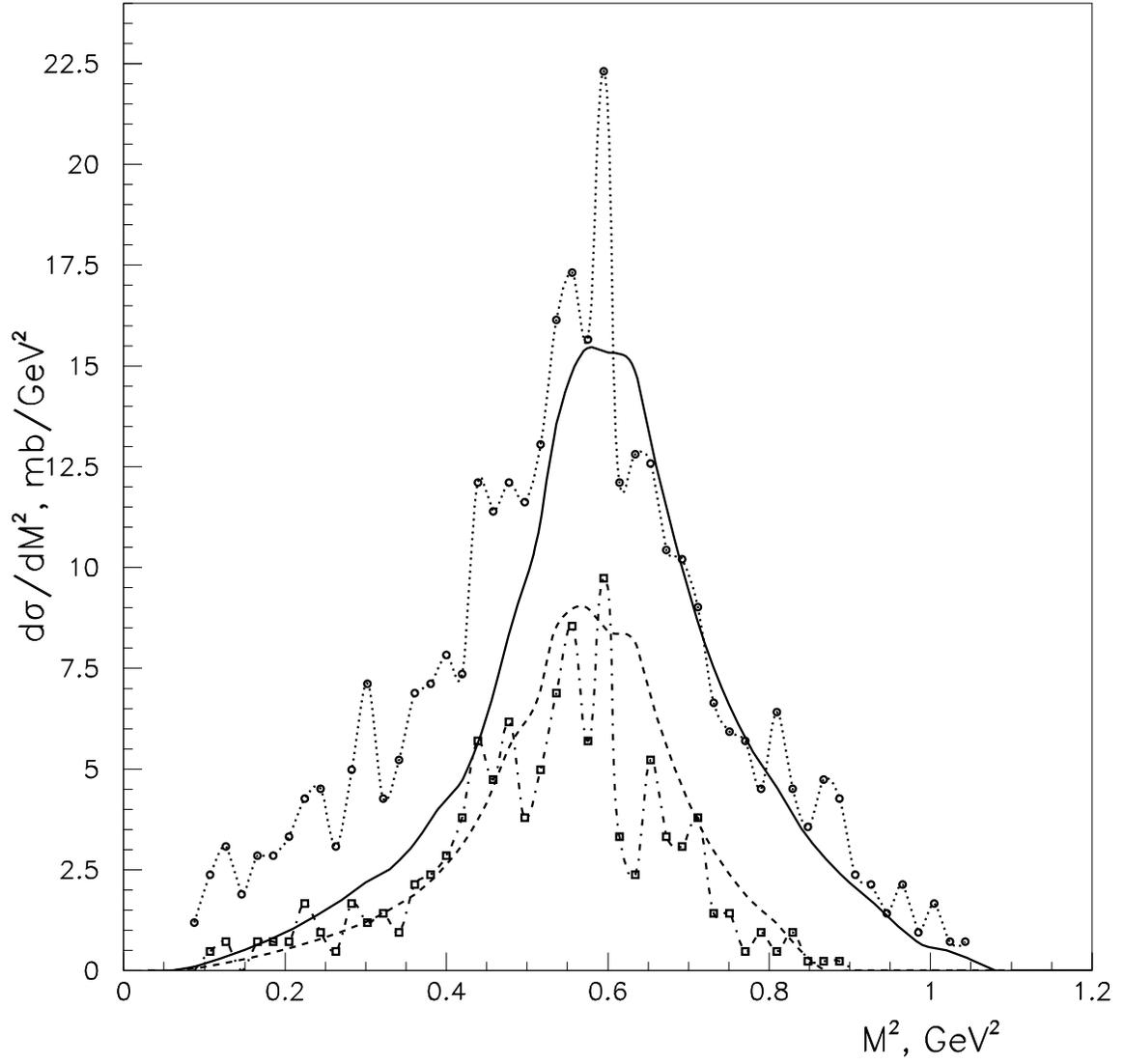,width=17cm,
            bbllx=0,bblly=0,bburx=567,bbury=841}}
\caption [] {Mass-squared distribution of 
$\pi^{+} \pi^{-}$ system in the reaction 
$\pi^{-}p \rightarrow \pi^{+} \pi^{-} n$ [14]
at $p = 1.59~GeV/c$, open circles correspond
to full range of $t$, open squares are
for limited $t$ (see text), 
the solid and dashed curves are our calculation.}
\label{diagram8}
\end{center}
\end{figure}

\clearpage
\begin{figure}[bhtp]
\begin{center}
\vspace*{-3.0cm}
\mbox{\epsfig
{figure=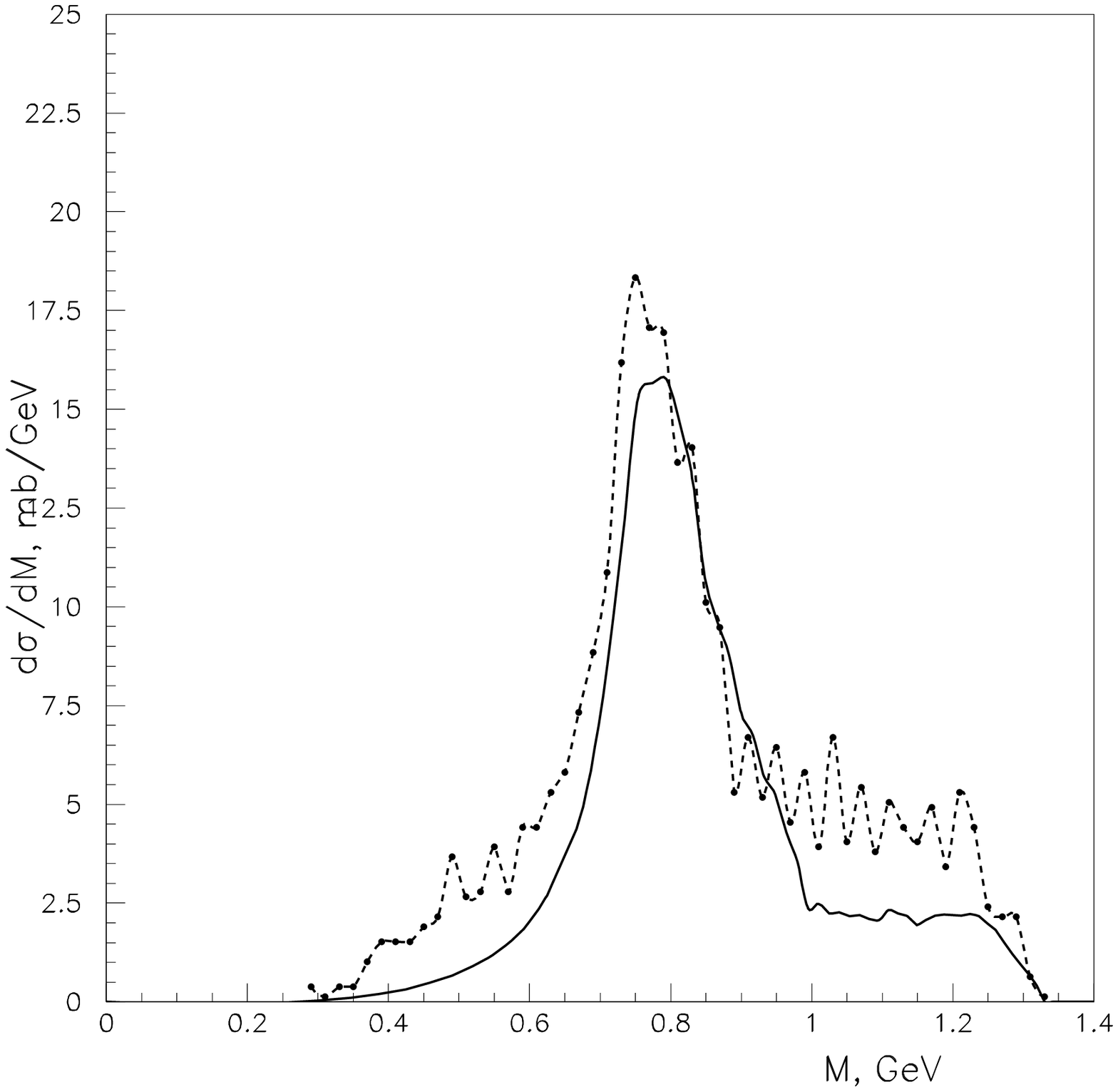,width=17cm,
            bbllx=0,bblly=0,bburx=567,bbury=841}}
\caption [] {Mass distribution of 
$\pi^{+} \pi^{-}$ system in the reaction 
$\pi^{-}p \rightarrow \pi^{+} \pi^{-} n$ [15]
at $p = 2.26~GeV/c$.
The solid curve is our calculation.}
\label{diagram9}
\end{center}
\end{figure}
\clearpage

\begin{figure}[bhtp]
\begin{center}
\vspace*{-3.0cm}
\mbox{\epsfig
{figure=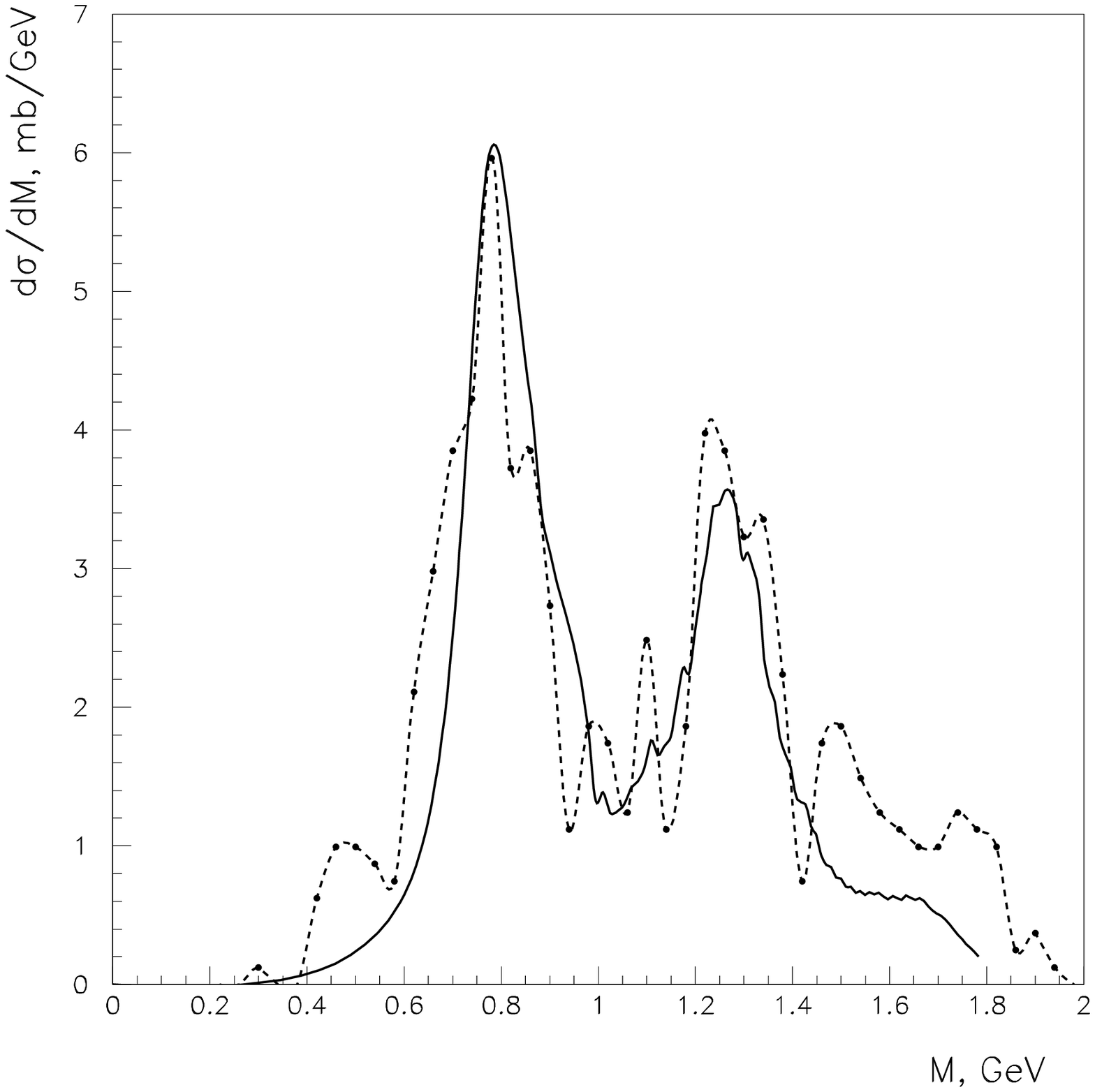,width=17cm,
            bbllx=0,bblly=0,bburx=567,bbury=841}}
\caption [] {Mass distribution of 
$\pi^{+} \pi^{-}$ system in the reaction 
$\pi^{-}p \rightarrow \pi^{+} \pi^{-} n$ [16]
at $ p = 4~GeV/c$.
The solid curve is our calculation.}
\label{diagram10}
\end{center}
\end{figure}
\clearpage

\begin{figure}[bhtp]
\begin{center}
\vspace*{-3.0cm}
\mbox{\epsfig
{figure=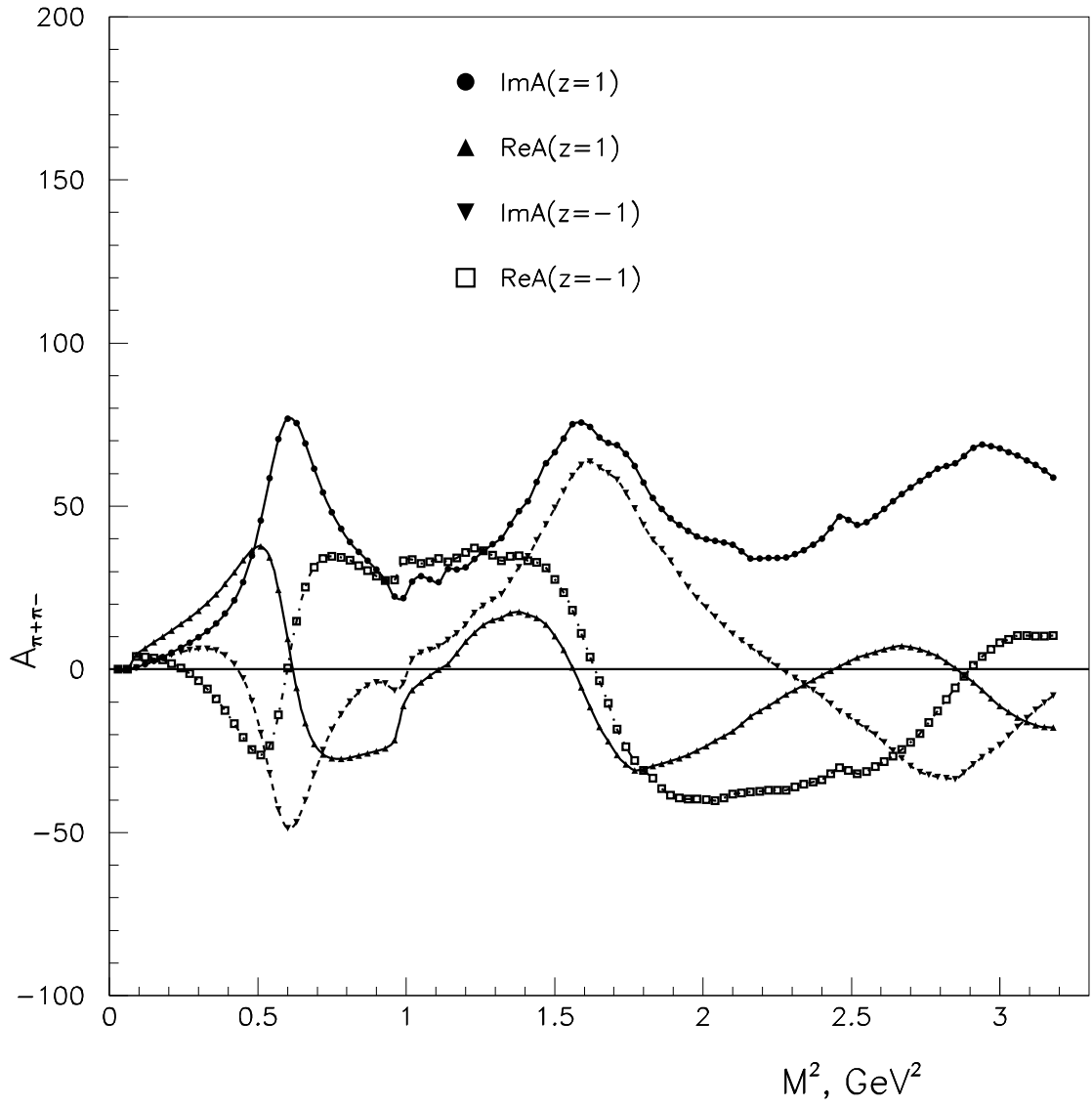,width=17cm,
            bbllx=0,bblly=0,bburx=567,bbury=841}}
\caption [] {Invariant $\pi^{+} \pi^{-}$  scattering amplitudes
within OPE model.}
\label{diagram11}
\end{center}
\end{figure}

\clearpage
\begin{figure}[bhtp]
\begin{center}
\vspace*{-3.0cm}
\mbox{\epsfig
{figure=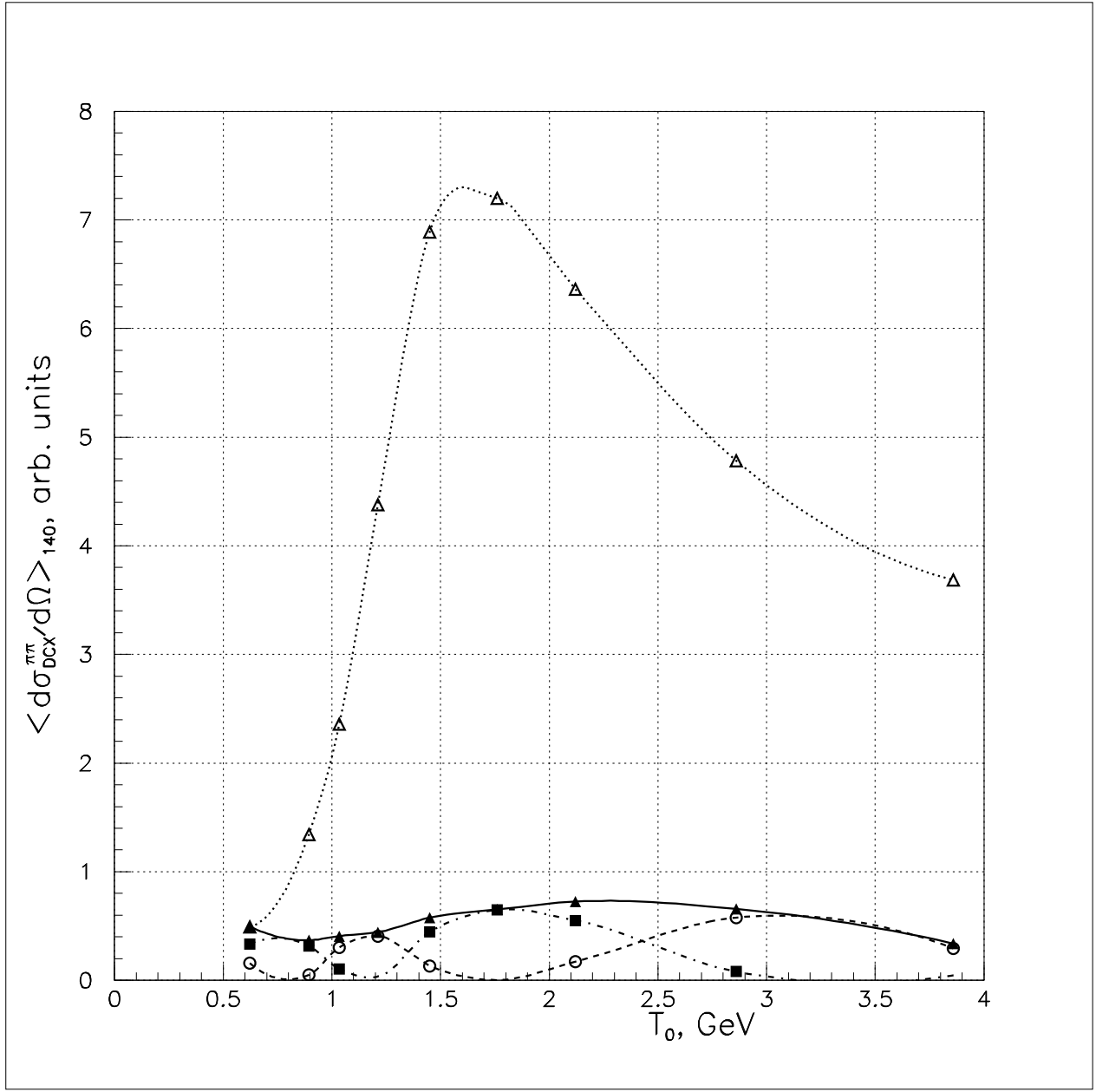,width=17cm,
            bbllx=0,bblly=0,bburx=567,bbury=841}}
\caption [] {$\langle d\sigma^
{\pi \pi}_{DCX}/d\Omega \rangle_{140}$ calculated
using Im$A(0)$ in Eq.(7) (empty triangles),
Im$A(z=-1)$ (full triangles), Re$A(z=-1)$ (empty
circles), and both Re$A(z=-1)$ and Im$(z=-1)$ (full
squares).} 
\label{diagram12}
\end{center}
\end{figure}

\clearpage
\begin{figure}[bhtp]
\begin{center}
\vspace*{-3.0cm}
\mbox{\epsfig
{figure=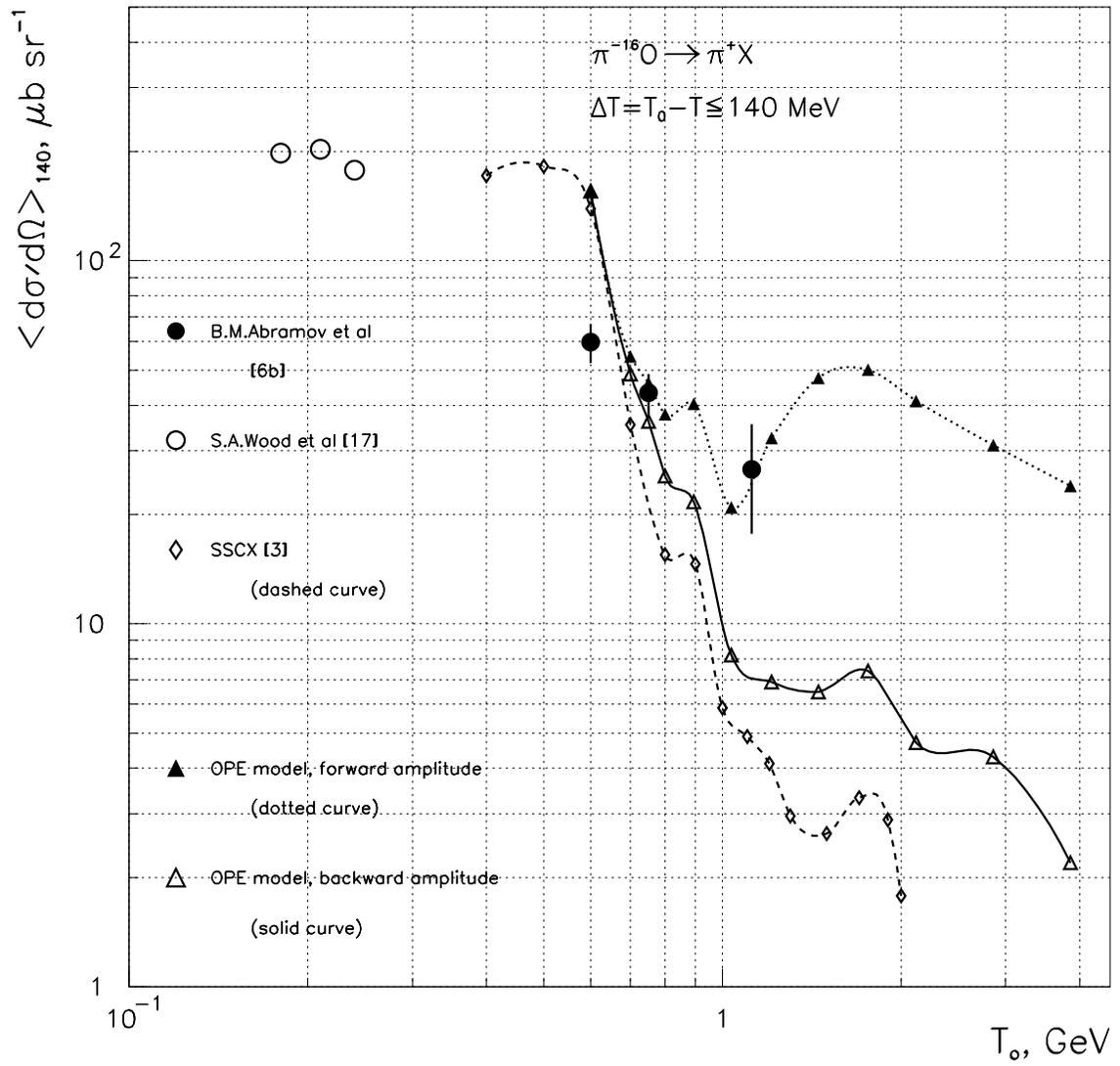,width=17cm,
            bbllx=0,bblly=0,bburx=567,bbury=841}}
\caption [] {$\sigma_{DCX}$ integrated over
$\Delta T$ range from 0 to 140 GeV.}
\label{diagram13}
\end{center}
\end{figure}

\end{document}